# Dynamic light scattering study on phase separation of a protein-water mixture: Application on cold cataract development in the ocular lens

V. Petta,[1,2] N. Pharmakakis,[3] G. N. Papatheodorou,[1,2] S. N. Yannopoulos[1,]*

[1] *Foundation for Research and Technology Hellas – Institute of Chemical Engineering and High Temperature Chemical Processes (FORTH / ICE-HT), P.O. Box 1414, GR–26504, Patras, Greece*

[2] *Department of Chemical Engineering, University of Patras, GR–26504, Patras, Greece*

[3] *Medical School, Department of Ophthalmology, University of Patras, GR–26504, Patras, Greece*

**Abstract**

We present a detailed dynamic light scattering study on the phase separation in the ocular lens emerging during cold cataract development. Cold cataract is a phase separation effect that proceeds via spinodal decomposition of the lens cytoplasm with cooling. Intensity auto-correlation functions of the lens protein content are analyzed with the aid of two methods providing information on the populations and dynamics of the scattering elements associated with cold cataract. It is found that the temperature dependence of many measurable parameters changes appreciably at the characteristic temperature ~16$\pm$1 $^o$C which is associated with the onset of cold cataract. Extending the temperature range of this work to previously inaccessible regimes, i.e. well below the phase separation or coexistence curve at $T_{cc}$, we have been able to accurately determine the temperature dependence of the collective and self-diffusion coefficient of proteins near the spinodal. The analysis showed that the dynamics of proteins bears some resemblance to the dynamics of structural glasses where the apparent activation energy for particle diffusion increases below $T_{cc}$ indicating a highly cooperative motion. Application of ideas developed for studying the critical dynamics of binary protein/solvent mixtures, as well as the use of a modified Arrhenius equation, enabled us to estimate the spinodal temperature $T_{sp}$ of the lens nucleus. The applicability of dynamic light scattering as a non-invasive, early-diagnostic tool for ocular diseases is also demonstrated in the light of the findings of the present paper.

---

* Corresponding author. E-mail: sny@iceht.forth.gr.

# I. INTRODUCTION

Protein condensation is of major importance in biosciences because condensed proteins can be either the cause or the corresponding early symptom of eye lens cataract or other neurodegenerative diseases, such as Parkinson's and Alzheimer's [1]. Cataract is one of the most frequently emerging diseases in the human kind being considered today as the most important cause of blindness worldwide. On general grounds, cataract is any opacification of the lens, which interferes with visual function producing increased light scattering. Lens protein aggregation and/or phase separation of the lens cytoplasm into protein-rich and protein-poor liquid phases have been considered as possible sources of cataract formation [2]. The lens is composed of about one thousand transparent, protein containing fiber cells. Its superior transparency is the result of the loss of organelles during lens evolution as well as due to the specific short-range structural ordering of the constituent proteins [3, 4], called crystallins.

Actually, the lens can be considered as a dense aqueous dispersion of crystallins whose concentration ranges from ~200 mg ml$^{-1}$ in lens periphery (cortex) up to ~500 mg ml$^{-1}$ in the lens nucleus. These values, especially the nucleus protein concentration, differ considerably among various vertebrates. It is therefore quite peculiar why such a dense dispersion is highly transparent. It was firstly proposed by Trokel [3(a)] that the high concentration of proteins in the lens has some degree of local order forming a "paracrystalline" state. Later, Benedek [3(b)] speculated on how a dense, amorphous packing of the proteins could lead to the suppression of the scattered intensity. More insight into lens transparency was provided by a small-angle x-ray scattering study [4] performed both in intact lenses and in lens homogenates (cytoplasmic extracts) over a wide range of concentration (3 to 510 mg ml$^{-1}$). This study revealed close similarities of the short-range spatial order of lens proteins with dense liquids and glasses.

Lens transparency is important in its own right for obvious reasons; structural studies, as described above, have illuminated many aspects. On the other hand, the loss of the transparency (cataract) is also significant since it engenders loss of visual function. Here one might argue that transparency loss might originate from conformational modifications of crystallins. However, studies on protein dispersions over a wide concentration range including transparent and non-transparent phases revealed that no conformational modifications of crystallins when cataract develops [4]. Changes of density and or concentration fluctuations have been recognized as the cause of cataract [2]. The dynamics of these fluctuations are best studied with the aid of dynamic light scattering (DLS) due to the matching of light wavelengths with the spatial dimensions of fluctuations. The importance of such studies lies primarily on the fact that it can provide easily measured spectral parameters, which could act as sensitive indicators of cataract onset. In other words, a reliable methodology for *non-invasive* and *early diagnosis* of ocular diseases and in particular cataract could be established.

An easily realizable model of nuclear cataract in the laboratory is the so-called *cold cataract* [5]. It can be easily induced by cooling the lens to temperatures below some characteristic or cold "cataract" temperature $T_{cc}$ located below the physiological one. $T_{cc}$ varies with the age and species of animal. Two main advantages of cold cataract over other types of cataract can be mentioned: (a) the effect is reversible; complete transparency can be achieved by warming the lens back to body temperature. (b) Cold cataract is induced gradually; this has the great advantage of offering a continuous monitoring of its stages due to the easy control of the external stimulus imposed, i.e. temperature. One is therefore able to examine the gradual appearance of the protein condensation process and thus to follow in detail the systematic changes of various spectral parameters of scattered light.

A DLS study of cold cataract in the calf lens has been presented elsewhere [6(a)], which is actually the sole DLS work reported dealing with cold cataract in the intact lens. It was found an increase in the size of the existing scattering elements as the temperature is lowered toward $T_{cc}$. That work aimed at elucidating the molecular mechanism responsible for increased light scattering from cataracteous lenses and judging between the two main theories of cold cataract: (i) aggregation or "particle" size increase with the formation of high molecular weight aggregates [7] and (ii) phase separation of the lens cytoplasm – into protein-rich and protein-poor phases – related with the onset of an increasing correlation length of concentration fluctuations approaching $T_{cc}$ [8]. The authors [6(a)] concluded that their data support the phase separation scenario. However, as it will become clear later from the results of the present study (section V.F.) no unambiguous evidence was provided in [6(a)] for the prevalence of phase separation in cold cataract due to misinterpretation of the various relaxation modes.

Although there is strong interest of studying morphological changes of proteins during cataractogenesis, DLS studies of cold cataract formation in intact lenses are largely missing, despite that this cataract model is an easily realized nuclear cataract model. Moreover, the advent of advanced multiple-tau correlators has provided the possibility for a more accurate determination of the complex relaxation functions of the eye lens that spans a wide time scale covering more that eight decays in time. In view of the aforementioned, we have undertaken – for the first time – a detailed DLS study of the cold cataract effect in porcine lenses over a wide temperature range from above to well-below $T_{cc}$. A major problem in diagnosing cataract by means of DLS is the lack of a systematic method of data analysis that results in the absence of a reliable parameter for cataract index. This paper concentrates also on this issue aimed at providing a more elaborate framework of data analysis in light scattering of ocular tissues.

The paper is structured as follows. Section II contains the experimental information including material preparation and DLS apparatus details. The data analysis procedures are presented in Sec. III, while the obtained results are contained in Sec. IV. Section V contains the discussion of the findings of this work and is divided, for clarity, in several subsections. It starts with a brief survey on experimental studies of the cold cataract effect. Then we discuss the results of this work in the framework of existing approaches on phase separation in the eye lens. Particular attention is paid to the temperature dependence of parameters describing the correlation functions and in particular to the fact that most of these parameters exhibit significant changes at some characteristic temperature identified with the cold cataract onset. Based on these findings, we also attempt an appraisal of the potentiality of the use of DLS as a non-invasive tool for diagnostic purpose. Finally, the most important conclusions drawn from the present study are summarized in Sec. VI.

## II. EXPERIMENTAL

### A. Materials

Fresh young porcine eyeballs (few hours after animal's death) were obtained from a local slaughterhouse under sterile conditions. Eyeballs, from which lenses were carefully excised, were stored at low temperatures (–5 $^{o}$C) and used for light scattering experiments during the first two days. The intact lenses were immersed either into a sterile physiological balanced salt solution (BSS) or into a hydrophobic fluid (silicon oil). Experiments showed that lenses in BSS, apart from nuclear cataract at low temperatures, were also developing cortical cataract of osmotic origin due to water exchange between the lens and the buffer solution. Cortical opacification deteriorates the laser beam transmission and prevents from obtaining reliable results from the nuclear part of the lens. On the contrary, the silicon oil revealed a more protective alternative for the lens (no penetration was possible) and moreover acted as a refractive index matching medium reducing to a large extent elastically scattered light from the silicon oil/lens interface. Therefore, only experimental results obtained from oil-immersed lenses will be presented and analyzed in this paper. The lenses were placed into a cylindrical pyrex tube of optical quality and appropriate diameter in order to host the porcine lens. The lens temperature was adjusted with high accuracy (0.1 $^{o}$C) by a temperature controller. Controlled cold cataract formation was achieved by gradually lowering the temperature starting from the physiological one (37 $^{o}$C). Equilibration of at least 30 min was taking place at each temperature before data collection. The results of three lenses taken at different times are analyzed and discussed in this work.

### B. Dynamic Light Scattering apparatus

DLS measures the fluctuations in light intensity scattered by a medium irradiated by a laser. In this work DLS provided the normalized intensity time auto-correlation function $g^{(2)}(q,t) = \langle I(q,t)\,I(q,0)\rangle / \langle I(q,0)\rangle^2$ at right angle ($\theta = 90^{o}$) over a broad time scale (nine decades) with the aid of a full multiple tau digital correlator ALV-5000/E supplemented by a fast unit that allowed to record $g^{(2)}(q,t)$ at short times (12.5 ns). The angular brackets indicate a time average. The scattering wavevector $q = 4\pi n \sin(\theta/2)/\lambda_0$ depends on the scattering angle $\theta$, the laser wavelength $\lambda_0$, and the refractive index of the medium $n$. The 488 nm laser line emerging from an Ar$^{+}$ ion laser was used with power on the lens of about 10 mW. The scattering geometry was arranged in such a way so as to record the intensity auto-correlation functions from the nucleus of the lens where cold cataract takes place. The lens was placed with its optical axis perpendicular to the scattering plane and the laser bean was directed along the diameter of the lens at the equatorial level. The scattered light was collected by a photomultiplier and transferred to the correlator for analysis after passing through either two pinholes (diameter 200 μm) or a single mode optical fiber.

Under certain conditions, such as Gaussian distribution of scattered light $g^{(2)}(q,t)$ is related to the desired electric-field time auto-correlation function $g^{(1)}(q,t)$ through the Siegert relation [9],

$$g^{(2)}(q,t) = B\,[1 + f^*|g^{(1)}(q,t)|^2]  \qquad(1)$$

where $B$ describes the long delay time behavior of $g^{(2)}(q, t)$ and $f^*$ is a coherence factor depending on instrumental parameters (pinhole sizes and distance from the detector, etc.) obtained experimentally from measurements of a dilute polystyrene/toluene solution. In our case, $f^* \approx 0.8$ for the pinhole arrangements and $f^* \approx 0.95$ for the optical fiber detection.

### III. DATA ANALYSIS

Figure 1 illustrates representative time correlation functions for two of the three lenses studied in this work. Since the physical characteristics of the scattering medium are associated with $g^{(1)}(q, t)$ the spectrum of the relaxation times of this function must be computed by solving Eq. (1). This is generally achieved in two ways. The first one employs the fitting of $g^{(1)}(q, t)$ with a discrete set of exponential or stretched exponential (SE) functions, as described below,

$$g^{(1)}(q,t) = \sum_i A_i \exp\{-[t/\tau_i(q)]^{\beta_i}\}, \qquad i: f, sl, usl \qquad(2)$$

where $A_i$ is the amplitude (zero-time intercept) of the $i^{th}$ decay step, and $\beta_i$ is the corresponding stretching exponent which is characteristic of the breadth of the distribution of the relaxation times and assumes values in the interval [0, 1]. For simplicity we drop the $q$-dependence in the following. The indices $f$, $sl$, and $usl$ denote the fast, slow, and ultraslow, individual decay processes, respectively whose characteristic times are shown by vertical dashed lines in Fig. 1. The selection of SE functions is dictated by the extremely broad range of the decay of the relaxation functions as shown in Fig. 1, which spans a range of over than eight decades in time and are particularly suited to account for the polydispersity of the probed particles.

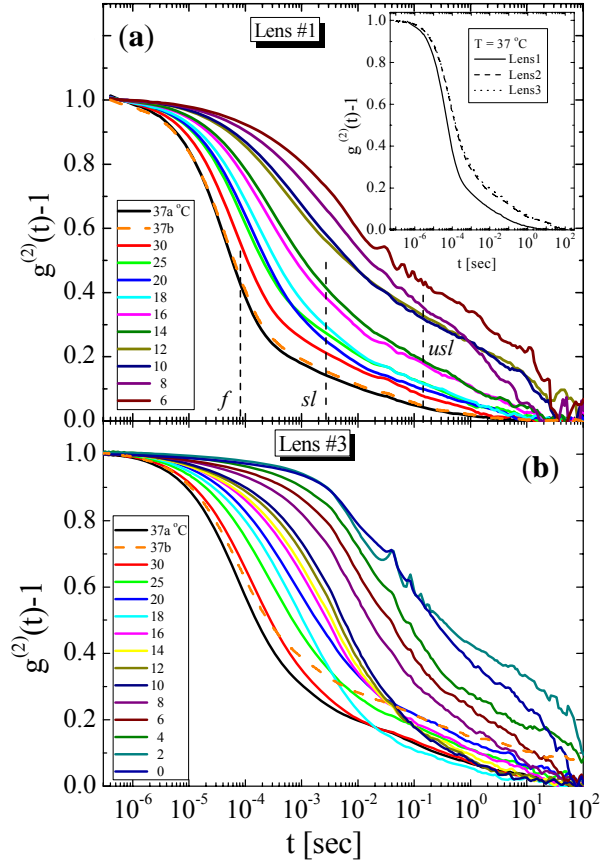

**Fig. 1:** *Normalized intensity auto-correlation functions of porcine lenses #1 (a) and #3 (b) for various temperatures as shown in the legend. The vertical dashed lines designate the fast, slow and ultraslow components revealed by the SE exponential analysis as described in the text. The correlation functions labeled 37b correspond to that measured after warming the lens to the physiological temperature after the cooling step. The inset shows the normalized intensity auto-correlation functions of the three lenses studied in this work at the physiological temperature (37 ºC).*

In an alternative procedure $g^{(1)}(t)$ is modeled as the sum of a large number of single exponential decays i.e.:

$$g^{(1)}(t) = \int L(\tau) \exp(-t/\tau) \, d\tau \qquad (3a)$$

which in the logarithmic representation of the relaxation time reads as:

$$g^{(1)}(t) = \int L(\ln \tau) \exp(-t/\tau) \, d\ln \tau \qquad (3b)$$

In this case, $L(\ln\tau) = \tau L(\tau)$ stands for the distribution of the relaxation times and its determination is usually accomplished with the aid of specific inverse Laplace transformation (ILT) techniques such as the CONTIN code [10]. The diffusion coefficient $D$ is determined by,

$$D = 1/\tau\, q^2 \qquad (4)$$

at the limit $q=0$; where $\tau$ is the relaxation time of $g^{(1)}(t)$. The apparent hydrodynamic radii of the protein "particles" are related to $D$ through the Stokes-Einstein relation,

$$R_h = \frac{k_B T}{6\pi\eta D} \qquad (5)$$

where $k_B$ is the Boltzmann constant and $\eta$ is the viscosity of the solvent. Analysis of time correlation functions recorded at various angles, at several temperatures, showed that, above $T_{cc}$, the relaxation processes of the two faster modes were found to be purely diffusive, i.e. $1/\tau$ was linear with $q^2$ as also was found in previous studies [6(a)]. The same conclusions could not be drawn for experiments performed below $T_{cc}$.

Intensity auto-correlation functions $g^{(2)}(t)$ were recorded with decreasing temperature starting from 37 °C, which is the physiological body temperature of mammals. The lowest temperature reached was 0 °C, where a considerable opacification of the lens occurred. In order to investigate the reversibility of the cold cataract, the lens was heated again to 37 °C and the experiment was repeated in order to check cold cataract reversibility.

## IV. RESULTS

Typical intensity time correlation functions (TCFs) are shown in Fig. 1 for two lenses. The curves have been normalized at short times in order to help revealing easily the changes brought about upon cooling. The inset shows a comparison between the intensity auto-correlation functions at 37 °C for the three different lenses studied in the present work. The comparison reveals some dissimilarities of the curves, especially at long times, associated with the presence of different fractions of high molecular weight assemblies (or aggregates) of proteins for the three lenses. This is reasonable since we deal with biological tissues taken from animals not controlled in the laboratory. The effect of enhanced scattering at long delay times on the qualitative and quantitative features of cold cataract will be discussed below. In brief, the results for lenses #2 and #3 were found practically identical.

The most evident change of the TCFs is a systematic increase of the contribution (amplitude) of the slow and ultraslow modes at the expense of the fast mode. Changes in the decay rate of the fast mode seem also to take place with cooling as is evident from the figure. In order to investigate the reversibility of the cold cataract, the lens was heated again to 37 $^{\circ}$C and the experiment was repeated showing negligible differences compared to the results of the first cooling step. The dashed curves (Fig. 1) show the physiological temperature correlation functions after warming the lens. Another important feature of the intensity auto-correlation functions is the appearance of ripples or oscillations with progressive cooling, which become more evident at low temperatures. Such ripples have not previously reported in DLS studies on cold cataract. Actually, periodic ripples in relaxation curves of gels obtained by DLS are rarely conveyed since they are considered abnormalities of the scattered light. Despite this, a wealth of information can be obtained from studies [11] on such oscillations as will be discussed in Sec. V.B.

The time dependence of scattered light (intensity trace) at constant temperature as well as the temperature dependence of the average intensity can also provide important information on the changes of dynamical properties of the density and concentration fluctuations. Figure 2 illustrates the temperature dependence of the average scattered intensity of lens #1. The inset shows the intensity traces of selected TCFs at various temperatures. It is evident a systematic increase of the average scattered intensity with decreasing temperature. The data suggest that this change takes place following two quasi-linear regimes, which cross at about 16 °C. It is also obvious (see inset) a change in the frequency and the amplitude of the fluctuations of the scattered light with decreasing temperature. Both effects are related to fluctuations of longer length scales (bigger "particles" or droplets of a new phase) that develop during cooling.

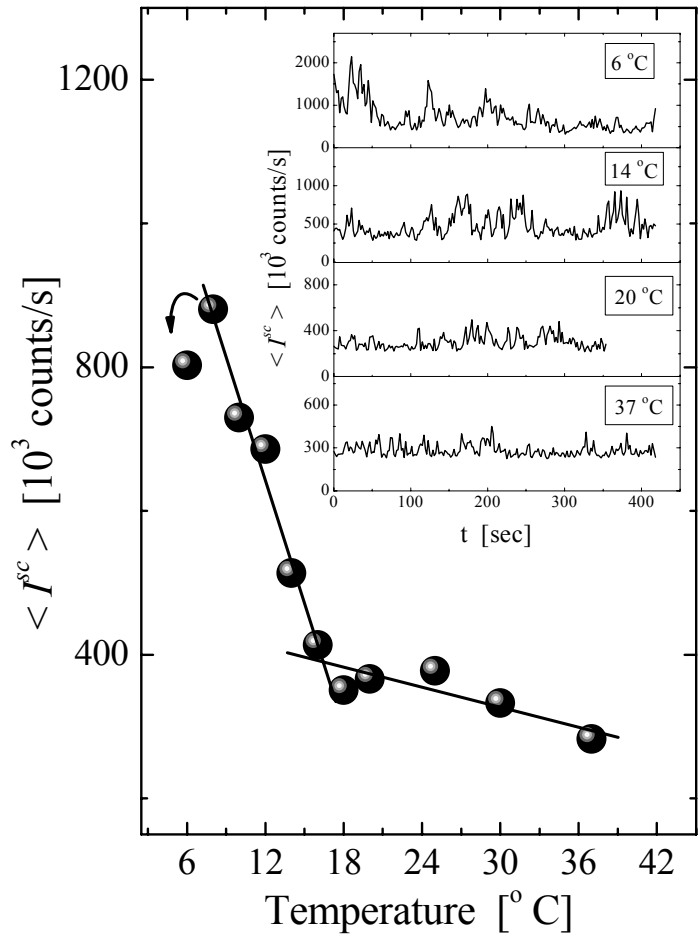

**Fig. 2:** *Temperature dependence of the average scattered intensity. Solid lines are drawn as guides to the eye to emphasize the two different scattering regimes crossing at ~16 °C. At the lowest temperature, lens turbidity causes a decrease in the light transmitted through the opaque lens to the detector. The inset shows the intensity traces (time dependence of the scattered intensity) at selected temperatures.*

To present the obtained results in a more quantitative way we have analyzed the experimental data using the two methodologies discussed in the previous section. The fact that the long time baseline is not uniquely determined could entail some uncertainty in the obtained results from the analyses methods we use. However, such uncertainties are expected to affect the very slow modes which are not the focus of the present paper. To affirm that the two fast modes remain unaffected by the analysis methods we have performed the CONTIN analysis on several correlation functions by selecting different fitting ranges, i.e. changing the long time limit of the fit. The results showed that the obtained distributions of the relaxation times at short times (fast and slow as denoted in the manuscript) were practically unaffected.

Representative fittings of the electric-field TCFs are shown in Fig. 3 for lens #1; similar data were obtained also for other lenses. Solid lines through the open symbols are best-fit curves obtained with the aid of Eq. 3. Equally good fits were achieved using the SE analysis (Eq. 2). The corresponding distributions of relaxation times resulted from the CONTIN analysis are also shown as dashed curves. The ILT distributions demonstrate the very complex relaxational nature of the porcine lens TCFs.

Indeed, four peaks in the decay time distributions are observed even at high temperature in the absence of any pathological condition of the eye. The first one corresponds to the fast decay mode that is characterized by the largest amplitude at 37 $^{o}$C. A very broad decay covering the time scale $10^{-3}$-$10^{1}$ sec follows that involves three distinct relaxation modes. Upon cooling, we observe a moderate shift of the four peaks of the ILT distribution to longer times, a broadening of the first peak, as well as a change in the relative intensities of the peaks, as well as the appearance of a new peak at very long times at low temperatures. The intensity of the first one exhibits a considerable decrease in relation to the other three modes. In addition, a non-zero baseline builds-up progressively indicating the onset of non-ergodicity of the scattering process [12].

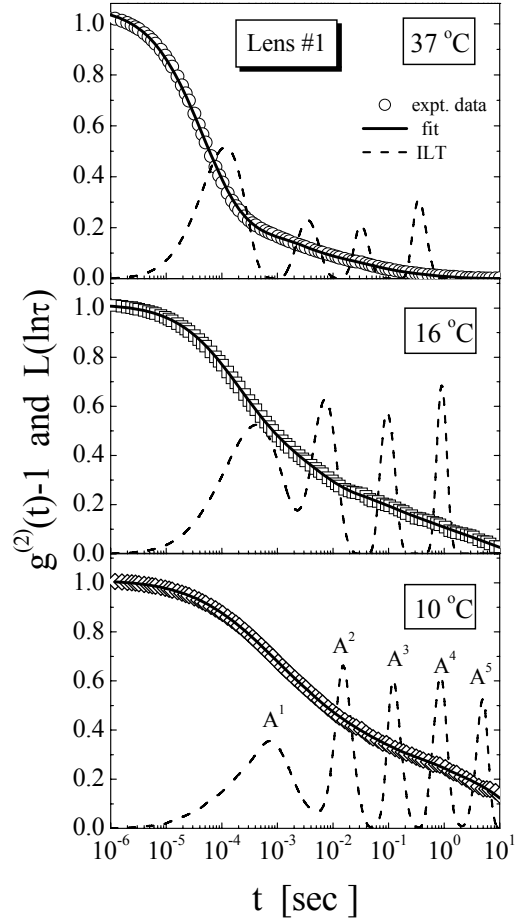

**Fig. 3:** *Representative fitting examples of the electric-field time correlation function $g^{(2)}(t)$ for lens #1 at various temperatures. Open symbols: experimental data. Solid line: best-fit curve with the aid of Eqs. 2 and 3, (best fits with both equations were indistinguishable). Dashed lines: distribution of relaxation times obtained with the aid of the ILT analysis (Eq. 3) using the CONTIN code.*

It should be mentioned at this point that the very slow dynamics developed in the lens nucleus upon cooling can be better studied by applying the relevant methodology suggested elsewhere [12] where the ensemble-averaged TCF is preferred over the time-averaged TCF. However, it should be noted that performing DLS studies on lens nucleus that involve the technique of ensemble averaging is not a realistic task. The lens is composed of fiber cells that contain the cytoplasm. Changing position of the scattering volume will complicate a lot the scattered light because, as biological studies have shown, the protein concentration varies along the diameter on the equatorial level of the lens as well as along the anterior-posterior optical axis. For this reason we have chosen to collect DLS data from the same scattering volume for all temperatures at which the experiment was performed. In any case, the data of the very slow modes studied in this paper are not considered on a quantitative ground.

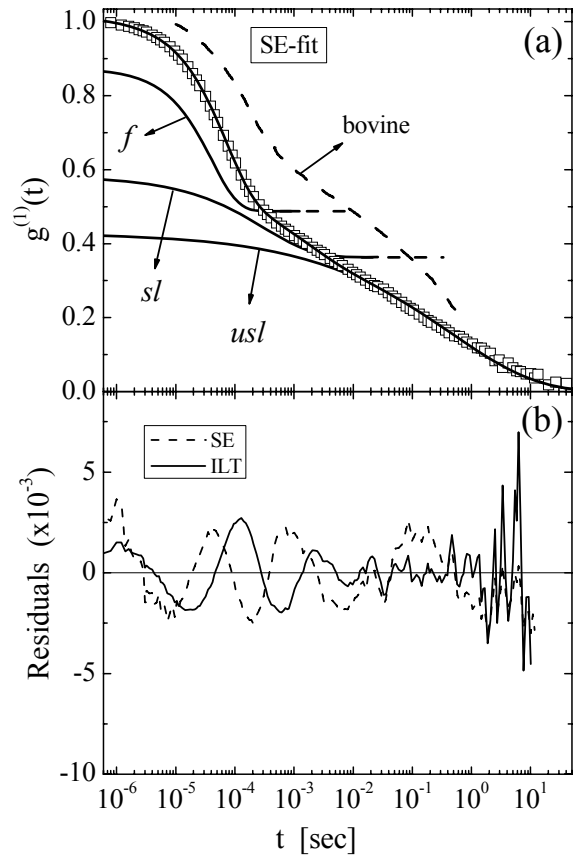

**Fig. 4:** *(a) A typical example of the SE analysis of the electric-field time correlation function at 37 °C where the three components of $g^{(1)}(t)$, i.e. f, sl, and usl are shown. For comparison, the dashed line corresponds to the correlation function of the calf lens at 14 °C arbitrarily scaled [taken from Ref. 6(a)]. (b) Fit residuals for the SE and ILT analysis methods.*

Up to now, most of the DLS data concerning dynamics of intact lenses were analyzed in terms of two exponential functions [6, 7]. This was mainly a problem of the limited function of old linear correlators that could operate at single, short sampling times. Measuring at different sampling time and joining the relaxation data it was possible to obtain a wider, in time domain, curve [6(a)]; [see dashed line in Fig. 4(a)]. However, the knowledge of the true baseline of $g^{(1)}(t)$ with high accuracy is of vital importance for the correct determination of both the decay times of the individual relaxation steps, as well as their relative amplitudes [9(b)]. Both are important in detecting early cataract stages since the decay times provide the diffusion constants or equivalently the hydrodynamic radii of "particles", and the amplitude ratio $A_{sl} / A_f$ [cf. Eq. 2] is used as the main indicator of cataract onset *in vitro* [6, 7] and *in vivo* [13] studies. However, it was shown [13(d)] that by using a logarithmic correlator, correlation functions of the rabbit lens recorded *in vivo*, and analyzed by various methods exhibited four relaxation processes, the origin of which will be discussed later.

To evaluate the validity of the CONTIN analysis presented above we provide hereafter a description of the data in terms of an exponential sum analysis since this was also the traditional method of data analysis in the vast majority of the previous works [6-8, 13]. We emphasize, however, that the limited time domain at which $g^{(2)}(t)$ was recorded in previous works restricted the analysis to the use of two simple exponential terms, usually termed as fast and slow. However, as described above by Eq. 2, we use stretched exponential functions that are considered to better describe the extremely broad decay of $g^{(2)}(t)$ accounting also for the polydispersity of the scattering elements. Figure 4 shows a representative example of fitting with the aid of Eq. 2 revealing an excellent fit of the data over a time scale covering 8 decades. The three individual decay steps fast, slow and ultraslow (*f*, *sl*, and *usl*) of $g^{(1)}(t)$ are also shown in the figure. An important outcome of the SE analysis is the temperature dependence of the various parameters. The lower panel in Fig. 4 illustrates a comparison between the residuals of the fit of Eqs. 2 and 3 to the experimental data, which demonstrates equally good fits for both methods of analysis. For comparison, we present in this figure the correlation function obtained [6(a)] in the DLS study of cold cataract of calf lenses. Although this curve covers a broad range of decay times revealing three types of scattering elements it is obvious, due to the lack of a baseline and a short-time plateau, that analysis of this incomplete data is still not capable of resulting in accurate conclusions about the relative intensity of each type of the involved scattering elements.

Figure 5 depicts the temperature variation of the exponents $\beta_i$ [cf. Eq. 2] expressing the non-exponential character of the relaxation functions, which is ultimately related to the polydispersity of the "particles" or the spatial heterogeneity protein environments. Details about the usefulness of these parameters as valuable indicators of the phase separation of the lens cytoplasm will be presented in Sec. V.D.

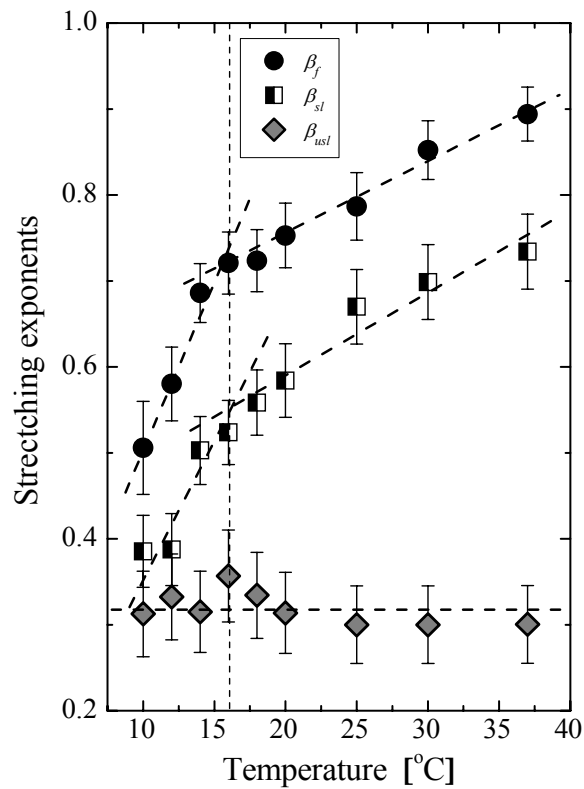

**Fig. 5:** *Temperature dependence of the stretching exponents of the three relaxation modes obtained by the SE analysis. Lines are used as guides to the eye. The vertical dashed line is used to indicate a change in the temperature dependence of the $\beta_f$ and $\beta_{sl}$; $\beta_{usl}$ is practically temperature insensitive.*

The parameters that were most studied in previous DLS studies of cataract are the relaxation times and the corresponding intensities or amplitudes of the scattering process related to the various types of particles. The former are related to diffusion coefficients and the latter are proportional to the populations of the scattering elements. The temperature dependence of the various ratios of the scattering intensities of the fast and slowly relaxing scatterers is shown in Fig. 6. The plot reveals that the intensity related to fast mode ($A_f$) diminishes systematically in relation to the total intensity assigned to all diffusion modes ($A_{tot}$), Fig. 6(a). This is evident from the results of both methods of analysis albeit with a different quantitative temperature dependence. At about 16 °C the decrease of the ratio becomes more prominent with cooling. Correspondingly, the scattering fractional intensity of the large slow mode ($A_{sl}/A_f$ and $A_{sl}/A_{tot}$) is depicted in Fig. 6(b) as obtained from the SE analysis.

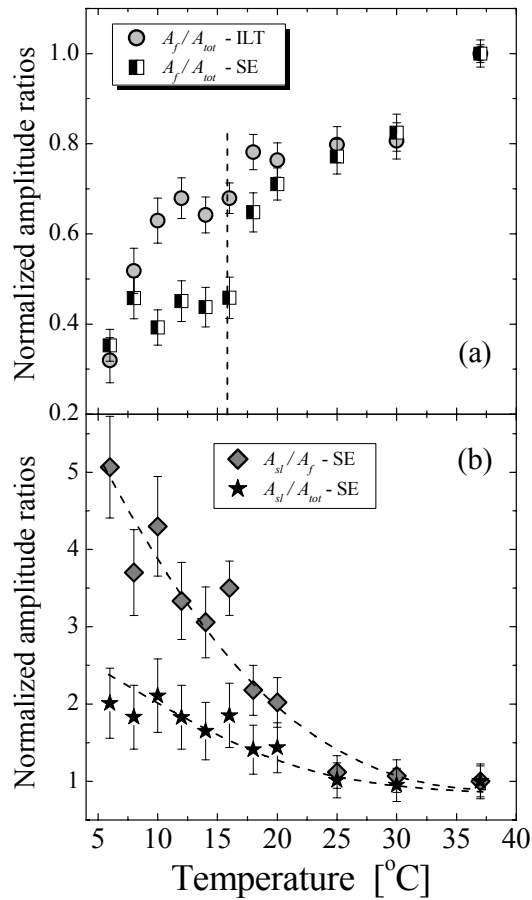

**Fig. 6:** *(a) Temperature dependence of the amplitude ratios of the fast ($A_f$) over the sum ($A_{tot}$) of the relaxation modes obtained by the SE and ILT analysis methods. (b) Temperature dependence of the amplitude ratios $A_{sl}/A_f$ and $A_{sl}/A_{tot}$ obtained by the SE analysis. The lines through the data points are used as guides to the eye.*

The changes of the relaxation times of the fast and slow process during cold cataract formation in porcine lenses as obtained from the SE and CONTIN analysis are shown in Fig. 7 in the form of an Arrhenius type diagram (i.e. logarithm of relaxation time vs. inverse absolute temperature). In the insets of Fig. 7 the same data are shown in linear coordinates. Average relaxation times are used which correspond to the most probable time of the ILT distribution and to the mean $<\tau_i>$ obtained from the $\tau_i^*$ parameters of SE analysis through the relation:

$$<\tau_i> = \tau_i * \beta_i^{-1}\Gamma(\beta_i^{-1}) \qquad (6)$$

where $\Gamma(x)$ is the Gamma function. The above relation provides a correction to the characteristic decay time taking into account the stretching (departure of $\beta_i$ from unity) of $g^{(1)}(t)$. As it is obvious for both fast and slow processes (see insets of Fig. 7), the decrease of temperature causes mild changes in the dynamics of protein diffusivity up to some characteristic temperature below which a severe slowing-down of protein motion takes place. The dashed lines, passing through the experimental data, represent best linear fits based on an Arrhenius-type relation between relaxation time and temperature, i.e.

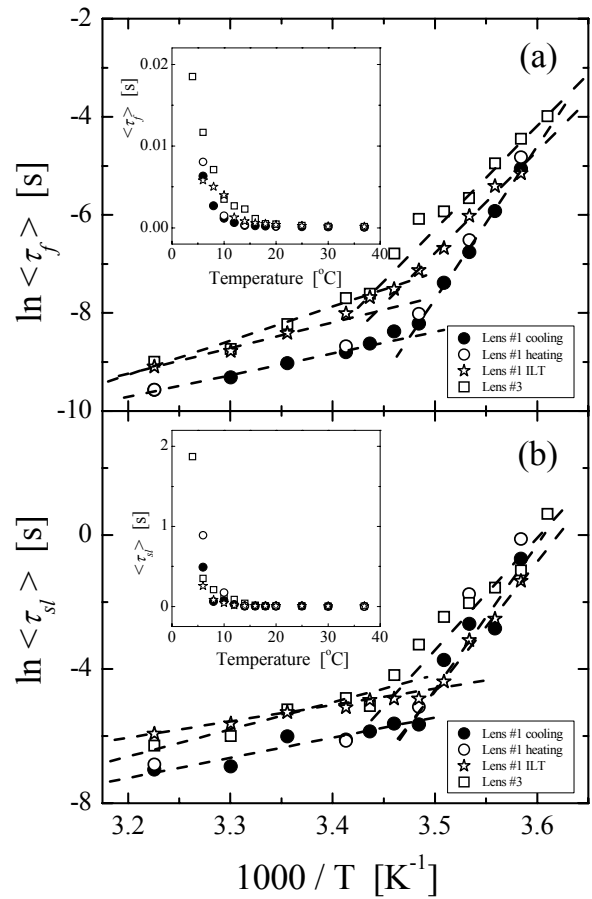

**Fig. 7:** *Arrhenius-type diagrams of the temperature dependence of the relaxation times of the fast (a) and slow (b) processes obtained by the SE and ILT analyses for lens #1 and #3. Dashed lines represent best linear fits to the experimental data. The insets show the temperature dependence of the relaxation times in linear temperature scale. Closed circles correspond to the measurements taken during the first cooling cycle of the experiment, while open circles represent the characteristic times obtained during the second cooling cycle for lens #1.*

$$<\tau> = \tau_0 \exp(E_a/RT) \qquad (7)$$

where the pre-exponential factor $\tau_0$ is a constant, $E_a$ is the apparent activation energy, and $R$ is the universal gas constant. For both lenses and for both processes (fast and slow) the critical point $T^*$ at which we have a slope change – or a change in the activation energy – occurs approximately at the same temperature. The obtained values for $E_a$ and $T^*$ are compiled in Table I.

**Table I:** *Apparent activation energies (in kcal mol$^{-1}$ K$^{-1}$) of protein diffusion and characteristic temperatures of the phase separation. f and sl denote fast and slow processes, H and L denote high and low temperature regions.*

|  | $E_{a,f}^H$ | $E_{a,f}^L$ | $E_{a,sl}^H$ | $E_{a,sl}^L$ | $T_f^*$ (°C) | $T_{sl}^*$ (°C) |
|---|---|---|---|---|---|---|
| Lens #1 (SE) | 8.8±0.5 | 61.8±1.6 | 11.8±2.2 | 86.1±15.9 | 15.2±0.5 | 14.4±0.8 |
| Lens #1 (ILT) | 10.4±1.3 | 40.8±5.4 | 9.7±1.0 | 76.2±6.3 | 17.4±1.2 | 13.2±1.1 |
| Lens #3 (SE) | 13.9±2.2 | 42.8±3.35 | 16.1±5.8 | 66.2±15.9 | 17.6±1.1 | 15.2±0.7 |

## V. DISCUSSION

### A. Brief background on cold cataract studies

Although DLS studies on the cold cataract effect have been indeed limited there are a number of works dealing with this phenomenon from the biophysical, biochemical and spectroscopic viewpoints. A brief survey on the history of cold cataract will also facilitate the discussion of our results in terms of the two main mechanisms of cold cataract, i.e. protein aggregation and phase separation of the lens cytoplasm. Cold cataract has been under investigation for more than a hundred years (see Ref. 14 for older references). Studies by Bon [14] were focused on changes in α- and β-crystallin levels in the lens, but definite experimental results did not prove that such changes could account for the phenomenon. His main observations were the following. (a) Cold cataract appears only in lenses of young vertebrates and is reversible upon heating the lens. (b) The ratio α- / β-crystallin concentrations was found very different between young and adult eye lenses; hence cold cataract was assigned to the high value of this ratio in the young eye lens. (c) $T_{cc}$ was found to depend on the species and on the age of the investigated animal. It is very interesting to notice that Bon was the first to suggest that the reversible opacification of the lens during cooling could be related to a *phase separation* between two-fluid phases in analogy to what happens in binary mixtures that possess a critical mixing point. The issue of phase separation as a possible mechanism for cold cataract, as well as for other cataract models, was subsequently adopted by many researchers [8] without reference to Bon's suggestion.

More systematic studies of cold cataract in young rat lenses were put forward by Zigman and co-workers [5]. A soluble protein fraction (cold precipitable protein or CPP) was found to precipitate both in lenses and in aqueous lens extracts (lens homogenates) below 10 °C. Apart from confirming observation (a) mentioned above, other important findings of that work include the following: (i) The CPP contains representative species of the three major classes of non-aggregated α-, β-, and γ-crystalline, where the concentration of the latter is much higher than the other two. The relative fractions of α:β:γ in CPP were found to be 23:13:64. (ii) A concentration of at least 3 mg/ml is required to observe the cold cataract effect. (iii) Cold precipitation depends on the ionic strength of the dispersion and the pH. (iv) Cold cataract can only be demonstrated in the lenses of young animals due to the presence of γ-crystallins, which constitute the major portion of the soluble lens proteins.

Based upon the aforementioned, as well as other findings, it was suggested [5] that cold cataract could be related to the balance between hydrogen and "hydrophobic" bonding. The mechanism responsible was related to conformational changes of proteins, although more recent studies [4] of transparent and opaque dispersions revealed that no conformational modifications of crystallins take place in the cold cataract phenomenon. In particular, it was considered that amino acids with non-polar side chains contained in lens proteins could participate in "hydrophobic" bonding, i.e. interactions involving nonpolar amino acid side chains of proteins. According to this mechanism, high temperature would facilitate the "hydrophobic" bond formation. Equivalently "hydrophobic" bonds may be broken during cooling and the freed nonpolar side chains of amino acids could engender insolubility. On the other hand, hydrogen bonds become stronger with decreasing temperature, predominating over the "hydrophobic" bonding. The competition between these two types of bonding was considered to determine protein stability and solubility [5]. The enhanced insolubility of γ-crystallins at low temperature is due to their small size, which implies that there are more hydrophobic groups exposed.

Spectroscopic studies of cold cataract were mainly conducted with the aid of nuclear magnetic resonance (NMR) spectroscopy [15]. These studies focused on the changes of relaxational properties of bound and unbound water molecules to proteins. It was thus possible, indirectly, to track changes in protein-water interactions in the development of the cold cataract effect. NMR results provided supporting evidence for the special role of γ-crystallins on this effect. The role of inhibitors that are deliberately added to the lens in order to prevent cold cataract development was also investigated [15]. Earlier work [16] had shown that a large fraction of γ-crystallins contains a significant number of hydrophobic residues, which are responsible for enhanced insolubility with decreasing temperature. Based on that work, NMR studies suggested that the presence of an inhibitor (acrylamide) causes aggregation of crystallins resulting in a decrease of hydrophobicity and cold precipitation. Finally, Raman spectroscopy was also utilized in the study of cold cataract of lenses concluding that no significant protein conformations cause or even accompany this effect [17(a)] as was suggested elsewhere [5]. However, it was found that some amino-acid (tyrosine) residues of the proteins do undergo changes in their hydrogen "microenvironment" during cooling [17(b)].

Other sporadic cold cataract studies of various lens properties include investigations of the optical performance of the lens [18(a)] and the polarization properties of light scattering in fish lenses [18(b)]. In the former work it was found that the focal length profile before and after cold cataract was identical suggesting that only supramolecular changes accompany cold cataract. In [18(b)] studies of the partially irreversible cold cataract in fishes revealed that γ-crystallins are not solely responsible for this effect because their very high concentration in fish lenses would engender cold cataract at higher temperatures in fishes than in mammals, while the opposite is true. Several works have also appeared were the effect of added substances on the inhibition of the cold cataract of the lens cytoplasm has been investigated (for a review see Ref. [2]). In the context of the above information we discuss in more detail the experimental findings of the present work.

### B. Onset of ripples in TCFs

An important effect accompanying cold cataract is the development of ripples in the TCFs. Ripples appear frequently in gels as manifestations of standing displacement waves set up in the gel structure due to some external mechanical vibration [11(a)]. Models of correlation functions have been developed that are able to provide information for the elastic moduli of the gels that refer to the polymer matrix alone [11(b)] as if there was no water present. In other investigations it was suggested that external mechanical vibrations need no be the cause of ripples in gel materials [11(c)]. In that work the oscillatory behavior was explained invoking a spring-rotor model in which the molecular motions inside the gel were modeled as vibrations of spring having various frequencies. Alternatively, ripples may appear in any case where a special order develops in the studied medium such as in cases of crystallization of supercooled liquids when heated at temperatures above the glass transition temperature [19]. Other causes of ripples have also been proposed [20]. The gel modulus calculation based on ripples frequency is possible in certain cases [11(a), (b)]. It is interesting to notice that ripples appear at temperatures below ~16 $^o$C for all lenses studied in this work. Their amplitude grows with decreasing temperature while their frequency remains practically constant. In view of these observations, we suggest that ripples might be used as an *in vivo* indicator of the early stages of cataractogenesis; however, more detailed studies are needed to support this suggestion.

**C.     Identification of the diffusion modes. Do fast and slow modes correspond to small and large scattering elements?**

We turn now our attention to the assignment of the four relaxation modes of the TCFs as revealed by the ILT analysis. Four relaxation modes were only reported in an *in vivo* DLS study of lenses with the use of a logarithmic correlator that is indispensable for observing very slow modes [13(e)]. To understand the origin of the four modes we need to resort to the assignment of the corresponding relaxation modes of crystallins' dispersions at various concentrations [21].

The assignment has been recently revisited [21(h)] in a detailed DLS study, which is the sole work on whole lens homogenates, over a wide range of temperatures and concentrations ranging from very dilute to dense comparable to lens nucleus. The advantage of this work in studying whole lens homogenates is that the analogy of α:β:γ crystallins in the lens and hence the interactions between homologous and heterologous proteins are also maintained in the dispersion.

In brief, the fastest relaxation process originates from the *long-time, collective diffusion mode* of the proteins. This has been validated by experiments showing that the diffusion coefficient of this mode becomes smaller with increasing concentration [21(h)]. From detailed analysis and using information from existing biochemical studies the second and third mode were to identified with the *long-time, self-diffusion modes* of isolated α-crystallin "particles" and high molecular weight assemblies of α-crystallins (HMα) present in the lens nucleus. Both modes exhibit strong volume fraction dependence showing appreciable decrease with increasing volume fraction. The volume fraction dependence of the diffusion coefficients was found to be satisfactorily described by recent predictions by Tokuyama and co-workers [22] on colloids with soft character. The self-diffusion modes become active in DLS due to *size* and *optical* (or scattering-power) polydispersity of the relevant α and HMα "particles". The origin of the (fourth) slow mode is not obvious. This diffusion mode is reminiscent to the very slow structural relaxation modes observed in structural glasses termed as long-range density fluctuations [19]. It is thus reasonable to assume that in highly dense colloidal suspensions, approaching their glass transition, analogous modes may also exist.

It is of crucial importance to comment at this point on the assignment of fast and slow diffusing modes to small and large scattering elements, as is the norm in the literature [6(a)]. This assignment has created confusion and has misled many authors to adopt the intensity ratio of these two modes $A_{sl} / A_f$ as an indicator of cataract development. This assignment is incorrect for the following reasons: (a) the fast relaxation mode represents a collective diffusion process and hence it cannot be associated with an apparent "particle" size at any finite concentration far from $c=0$. (b) The slow mode cannot be associated with the large or aggregated scatterers but rather with the native α-crystallins as was demonstrate din [21(h)]. We will return to this point in the discussion of the molecular origin of cold cataract in Section V.F.

### D. Temperature dependence of the stretching exponent: a probe of the spatial heterogeneity and the onset of phase separation

The determination of the non-exponential parameters $\beta_i$ in Eq. 2 is an important element of the present study because in previous studies [6(a)] single exponential analysis was performed, thus overlooking an important piece of information. Two significant observations emerge from Fig. 5: (a) the very low magnitude of $\beta_{usl}$ which amounts to ~0.3, being also temperature insensitive; and (b) the strong temperature dependence of $\beta_f$ and $\beta_{sl}$ which exhibit parallel trends. The very low magnitude of $\beta_{usl}$ (extreme stretching) implies the existence of a broad distribution of relaxation times and thus justifies the finding of the CONTIN analysis that revealed at least two scattering elements distributions in the time scale $10^{-2}$–$10^{1}$ sec. The parameters $\beta_f$ and $\beta_{sl}$ systematically decrease with cooling exhibiting a faster rate of decrease below ~16 $^o$C, which is near the temperature where the mean scattered intensity exhibited a change in its growing rate (see Fig. 2).

According to the mode identification given in the previous subsection, $\beta_f$ and $\beta_{sl}$ account for the stretching exponent of the collective diffusion mode and the self-diffusion of the α-crystallins, respectively. The fact that $\beta_{sl}$ becomes appreciably lower than unity below $T^*$ points to the increased polydispersity of α-crystallins at temperatures below the cold cataract onset. The same reasoning cannot be applied for $\beta_f$ because the collective diffusion mode cannot strictly be associated with single particle dynamics. However, the increasing departure of the stretched exponential parameter in this case may indicate an enhancement of the *spatial heterogeneity,* or alternative the generation of local microenvironments with distinct dynamical properties as it commonly occurs in supercooled liquid approaching their glass transition temperature [19]. The development of spatial heterogeneity is indicative of the system's incipient phase separation with decreasing temperature that finally will engender lens opacification (cataract).

### E. Temperature dependence of diffusion coefficients: Estimation of the cold cataract and spinodal decomposition temperature

Relaxation times contain a wealth of information since important transport properties such as diffusion coefficients can be determined and hence apparent hydrodynamic particle radii and interparticle interactions can be assessed. The detailed analysis presented in the previous sections has allowed us to elucidate the interesting behavior of the dynamics of the collective and the self-diffusion processes as shown in Fig. 7. The apparent activation energy (or the slope of the Arrhenius curves) changes as a function of temperature exhibiting a strong increase with cooling. The straight lines represent linear fits of the limiting low- and high-temperature behavior and when extrapolated they cross at $T^*$, the values of which are listed in Table I. At the moment, the actual change of $E_a(T)$ is speculative, i.e. it is not clear if there are two regimes of constant activation energy or if this quantity follows a smooth temperature dependence with two limiting constant activation energies; though the former scenario seems more likely. The latter is reminiscent of the diffusive motions or structural relaxation of glass-forming supercooled liquids when the temperature is lowered towards the glass transition temperature [19, 23]. The increase of $E_a$ with decreasing temperature in supercooled liquids has been assigned to the onset of cooperativity in molecular motion. In particular, with decreasing temperature, molecular moves require the cooperative rearrangement of neighboring molecules and hence the apparent activation energy for a "diffusion step" requires overcoming a larger barrier in relation to what happens at high temperatures.

Considering the cold cataract as a phase separation phenomenon in protein-rich and protein-poor phases below $T_{cc}$ [8], the diffusion coefficient can provide apparent $R_h$ values (Eq. 5) at high temperatures, i.e. above the coexistence curve ($T > T_{cc}$). On the other hand, if the temperature of the lens is lowered below $T_{cc}$ so as to enter in the two-fluid phase the above description fails and thus diffusivity near the critical point is related through the same relation to the correlation length $\xi$ of critical fluctuations, i.e. $D = \dfrac{k_B T}{6\pi\eta\xi}$ [24]. The correlation length exhibits a strong growth (divergence) with decreasing the temperature. Figure 8 illustrates the apparent $R_h$ values and correlation lengths $\xi$ for the fast (a) and slow (b) modes. $R_h$ values obtained from the fast relaxation mode do not represent true hydrodynamic radii since this mode manifests collective diffusion, see Sec. V.F. for details. For comparison, the corresponding data of the small and large scatterers in the center of the intact nucleus of the bovine lens are shown in Fig. 8 taken from [6(a)] where it was possible to estimate particle size only for $T > T_{cc}$. The two data sets differ in magnitude due to either the difference between the two species (bovine and porcine lenses) and/or due to the higher errors in relaxation time estimation from $g^{(1)}(t)$ of Ref. [6(a)]. In the estimation of the apparent particle radii we have considered the viscosity change of water which between 37 °C and 4 °C differs by a factor of 2.

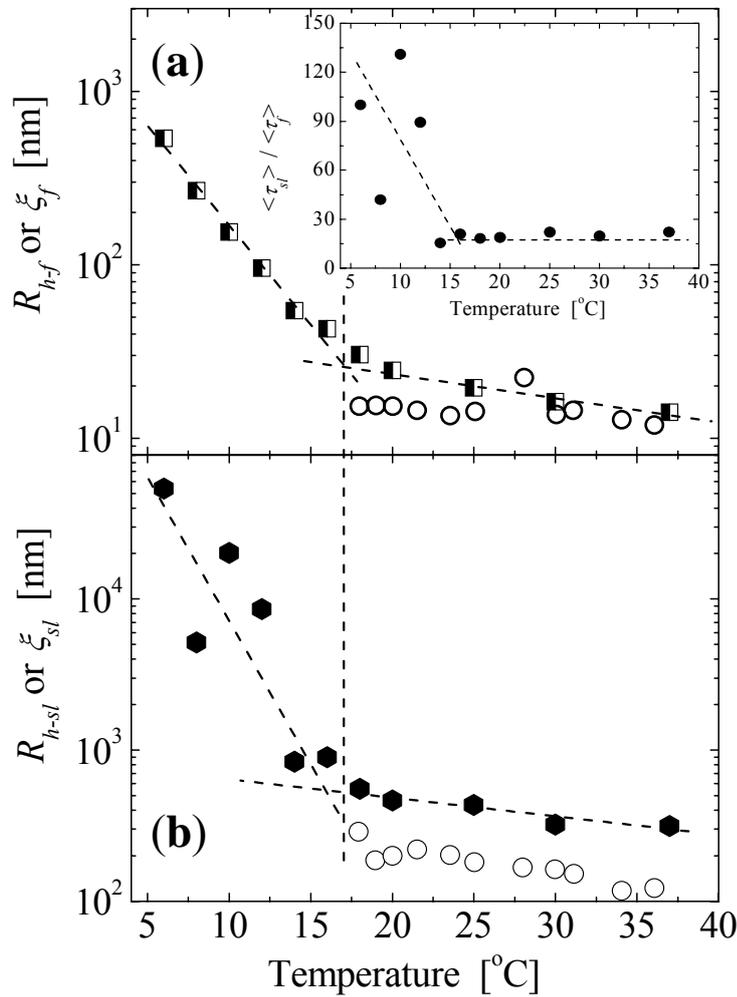

**Fig. 8:** *Temperature dependence of the hydrodynamic radii $R_h$ and correlation lengths $\xi$ for the small (a) and large (b) scattering elements. The vertical dashed line denotes $T_{cc}$. The open circles represent the corresponding data of the bovine intact lens nucleus form Ref. [6(a)]. The inset in (a) illustrates the ratio of the relaxation times related to small and large "particles"; see text for details.*

Figure 8 reveals again the interesting feature of a marked change in the temperature dependence of the apparent particle size at about 16±0.5 °C. It is also obvious a systematic increase of particle size when lowering the temperature from the physiological one towards $T_{cc}$, which implies a moderate particle aggregation. The ratio of the relaxation times $\tau_{sl} / \tau_f$ is practically temperature insensitive up to 16 °C (see inset of Fig. 8) with a mean value of ~18, while no systematic dependence exists at temperatures below the onset of the liquid-liquid phase transition. The dramatic change occurring in the protein/water mixture static (equilibrium) and dynamic properties below the phase separation temperature is responsible for the remarkable increase of the correlation length $\xi$ for both fast and slow relaxation modes. Obviously, the results in this temperature range are only qualitative.

The ever growing relaxation time or equivalently the continually decreasing diffusivity indicates that at a characteristic temperature "particle" motion will cease because the diffusion constant will vanish. This characteristic temperature is obviously located below $T_{cc}$, which corresponds to a point on the coexistence curve, and can be identified with a point $T_{sp}$ along the spinodal line located inside the co-existence curve. The part of the phase diagram between the co-existence and the spinodal curves denotes the [concentration, temperature] range in which the mixture can be supercooled (in case of an upper critical solution temperature phase diagram) without spontaneously separating into two fluid phases. In previous cold cataract studies [6(a)] in the calf and human lenses, diffusivity data from DLS measurements were limited to temperatures above $T_{cc}$, while in the present study we have been able to extend this range to well below $T_{cc}$. Based upon theories of binary mixtures the temperature dependence of the diffusion coefficient in terms of the spinodal decomposition temperature $T_{sp}$ and the corresponding critical exponent $\gamma$ [8(a)] can be written as:

$$D(T) = \alpha(T - T_{sp})^{\gamma} \qquad (8)$$

where $\alpha$ is a proportionality constant. Fittings of Eq. (8) to the experimental data for the fast and slow process are shown in Fig. 9. The obtained spinodal temperatures for the fast and slow diffusing modes are 5.1±0.5 °C and 4.8±0.8 °C, respectively. These values are much lower than the spinodal temperatures of the intact calf lens nucleus located at about 24 °C ; however, they are closer to the corresponding values of the intact human nucleus with $T_{sp}$ = -5 °C [8(a)]. This result is in accordance with the general perception that the porcine lens is a more realistic model for human lens than other mammal lenses. This fact indicates that the porcine lens cytoplasmic content behaves in cold cataract in a way intermediate between that of the very sensitive in temperature variations calf lens and that of the less sensitive human lens. It is also important to notice that the $T_{sp}$ values of the calf lens were obtained by linear extrapolation ($\gamma$ = 1) of diffusivity data at temperatures above $T_{cc}$.

We would like to comment on the fact that our temperature difference $\Delta T=T_{cc}-T_{sp}$, for the porcine lens nucleus (~18 °C), is appreciably larger than that estimated by Benedek et al. [8(a)] for the bovine lens nucleus (~3 °C). In that study [8(a)] $T_{cc}$ was identified with the temperature at which the lens scattered light intensely due to lens opacity. This is an arbitrary definition which depends upon the experimental resolution. Using this definition, $T_{cc}$ was located at ~25 °C for bovine lens nucleus [25]. Apart form the arbitrary determination of $T_{cc}$, as stated above, $T_{sp}$ was also determined with a large error in [8(a)] due to the linear extrapolation [$\gamma=1$ in Eq. (8)] and hence was considerably overestimated.

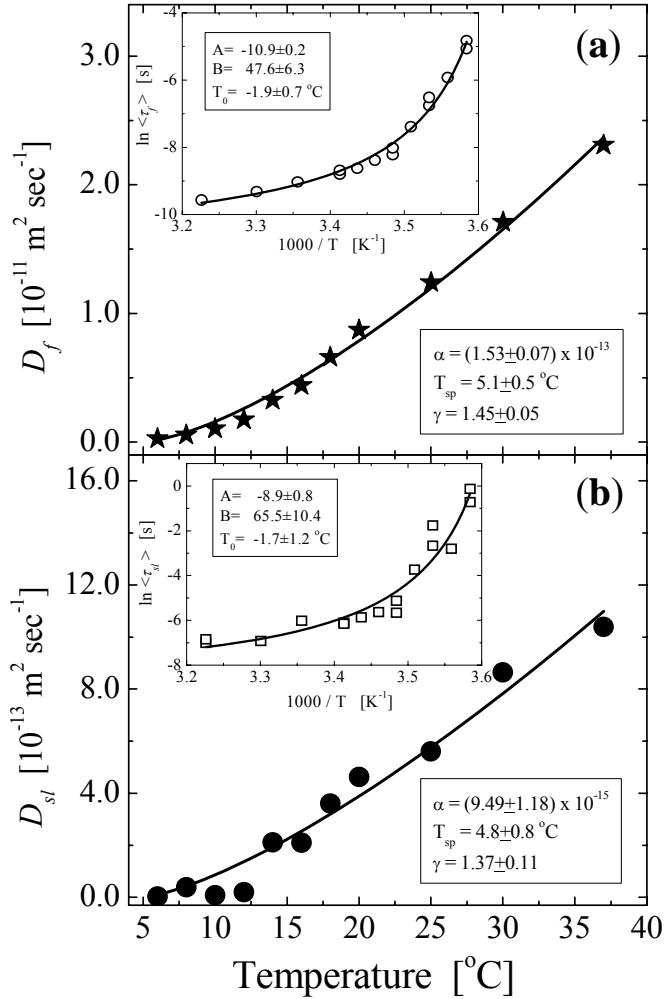

**Fig. 9:** *Temperature dependence of the diffusion coefficients for the fast (a) and slow (b) modes for lens #1. The solid lines represent fits with the aid of Eq. 8. The insets show the Arrhenius-type plots of the fast and slow modes fitted with the aid of the VFT relation (Eq. 9).*

As a result, the difference $\Delta T=T_{cc}-T_{sp}$ was found too small in [8(a)]. Our ability to record data at low temperatures allowed us to extrapolate the diffusivity temperature dependence, with higher accuracy, and thus to obtain a more reliable determination of $T_{sp}$ which yielded a larger value for $\Delta T=T_{cc}-T_{sp}$.

Following an approach common to glasses and supercooled liquids [23] the rapid decrease of diffusive motions at low temperatures can be modeled by a modified Arrhenius equation that contains a temperature dependence activation energy, i.e. the so-called Vogel–Fulcher–Tammann (VFT) equation,

$$<\tau(T)> = \tau_0 \exp\left(\frac{B}{T-T_0}\right) \qquad (9)$$

where the parameters $\tau_0$, $B$, and the "ideal glass transition temperature" $T_0$ where diffusivity vanishes are temperature independent quantities. $T_0$ of Eq. (9) is identified with $T_{sp}$ of Eq. (8). Fits to the above equation are quite satisfactory as shown in the insets of Fig. 9. The spinodal temperatures $T_0$ for the fast and slow diffusing processes are found at $-1.9\pm0.7$ °C and $-1.7\pm1.2$ °C, respectively. It is evident that fitting relaxation time data with Eq. (9) is preferable since diffusivity data approach zero at low temperatures exhibiting stronger scattering than the corresponding relaxation times. The spinodal temperatures of the VFT relation for the fast and slow decay modes are almost identical and lower than those values obtained by Eq. (8).

### F. On the origin of cold cataract: Aggregation vs. phase separation

One of the main objectives of the present paper is to provide new, properly analyzed data, which will help better understand the two main conjectures for cold cataract formation. Though previous DLS studies [6(a)] claimed to be consistent with the concept of phase separation below $T_{cc}$ the results of the present work clearly demonstrated that the criteria used in previous studies were misleading. Being unable to conduct DLS measurements below $T_{cc}$, Delaye *et al*. [6(a)] used data obtained only in the homogeneous, not-phase separated, regime. The criteria used to support phase separation were: (a) the existence of droplets of a new phase and (b) an increase in the apparent size of the droplets as the temperature is lowered towards $T_{cc}$. However, two problems appear here inspecting Fig. 8 in [6(a)]. First, "large" particles that were considered as droplets in [6(a)], already existed at the physiological temperature (37 $^{o}$C) and thus cannot be considered as new nucleating phase. Second, the increase in the apparent size of the large particles is moderate and actually it heralds the change if interparticle interactions instead of actual particle size increase. Determination of the latter in the dense lens cytoplasm via Eq. (5) equation is not possible.

In addition to the above facts on the misinterpretation of the observations (a) and (b) as indicative of phase separation, we would like to mention that according to the more detailed interpretation of the lens nucleus complex correlation function given in Section V.C. and in Ref. [21(h)] *the fast and slow modes do not correspond to the "small" and "large" scattering elements* invoked in Ref. [6(a)]. Instead, the fast mode is related to the collective diffusion process of the lens cytoplasm and the slow mode describes the long-time self-diffusion coefficient of α-crystallins, which represent the "small" particles of Ref. [6(a)]. Diffusion properties of "large" particles or high molecular weight assemblies of α-crystallins (HMα) are manifested through the third relaxation mode revealed by the ILT analysis. However, due to the large number of peaks in the ILT distribution functions, the area ratio $A^3 / A^2$ which reveals aggregation or droplet size increase cannot be determined with high accuracy and is only qualitative.

Clear indications of an incipient phase separation below $T_{cc}$ provided by the present study include: (a) the appearance, below $T_{cc}$, of a new very slow relaxation mode, see peak $A^5$ in the ILT analysis in Fig. 3 and more important (b) the temperature dependence of the stretching exponents for the fast and slow decays steps of the correlation functions. The value of the relaxation time of the new, very slow, mode justifies an apparent scattering element size of the order of few microns and hence can be envisaged as the manifestation of the nucleating droplets. The significant non-exponential form of the fast and slow processes, i.e. much lower value of $\beta_f$ and $\beta_{sl}$ than unity below $T_{cc}$, reflect the formation of spatial heterogeneity that reveals the creation of local microenvironments with distinct dynamical and hence compositional properties. The development of spatial heterogeneity is a direct consequence indicative of the system's incipient phase separation with decreasing temperature that finally will engender lens opacification (cataract).

### G. The nature of droplets in the phase separation regime

We close this section by a brief reference to the nature of the new phase droplets emerging from the liquid-liquid transition during cooling. The composition of these droplets has been a matter of discussion for many decades. The central role of γ-crystallins' concentration and in particular their attractive interactions have been recognized in several studies [8(f), (g)], although this view has been questioned in [18(b)]. Zigman *et al.* [5] found that the relative fractions of α:β:γ in droplets (or CPP) were found to be 23:13:64. More recently, the structure, distribution, and nature of the scattering elements associated with cold cataract formation in young rat lenses were studied in intact lenses using light and electron microscopy [26]. It was identified that cold cataract is accompanied by the formation of large spherical droplets (~1.5-10 μm in diameter) within all nuclear fiber cells and some deep cortical fiber cells as well. Each droplet was found to contain α-, β-, and γ-crystallins. Qualitative differences in droplet sizes were reported in cold cataract of lens homogenates and intact lenses; sizes were reported [26] larger in the latter owing to the presence of lens membranes.

## VI. SUMMARY AND CONCLUDING REMARKS

A detailed dynamic light scattering study of the cold cataract effect in intact porcine lenses *in vitro* has been undertaken. The study has been conducted for three different lenses in order to check the reproducibility of the results. The intensity autocorrelation functions were recorded from the nuclear part of the porcine lens at various stages of the cold cataract effect. The data were evaluated with the aid of two analysis methods, i.e. using stretched exponential functions and the inverse Laplace transformation in order to extract the parameters related to the populations and dynamics of the scattering elements. This combined analysis revealed the temperature dependence of many measurable parameters changes appreciable at the characteristic temperature ~16±1 $^{o}$C which is associated with the onset of cold cataract or phase separation of the lens cytoplasm.

Parameters that exhibit this change are the following. (i) The total (average) scattered intensity at fixed temperature. (ii) The stretching exponent of the fast and slow components of the TCFs related to the heterogeneity of the local environments of the proteins and the polydispersity of the scatterers, respectively. (iii) The ratio of the scattering intensities associated with the fast and the slow component, $A_{sl} / A_f$. (iv) The apparent activation energy of the fast and slow component when plotted in an Arrhenius type diagram. (v) The apparent hydrodynamic radii of the scattering elements. (vi) The ratio of the characteristic relaxation times $\tau_{sl} / \tau_f$. (vii) The appearance of oscillations (ripples) in the TCFs.

While from the viewpoint of basic science all the above observations are important parameters that give direct evidence for the onset of cold cataract, on practical grounds one cannot use many of them to make an early, non-invasive diagnosis of cataract at *in vivo* conditions. So far, cataract prediction is based on the relative intensity ratio between the slow and fast process of the intensity TCFs, i.e. $A_{sl} / A_f$ which has erroneously considered as indicative to the population of the corresponding small and large "particles". Our work showed that this ratio is smoothly growing with decreasing temperature. Although this factor seemed as a rather reasonable indicator of the formation of larger particles, in the spirit of the interpretation presented in Sec. V.C. it is now obvious that the $A_{sl} / A_f$ ratio cannot account for the aggregation process. Moreover, a single measurement can lead to elusive conclusions due to the strong temporal fluctuations of the scattered signal. Indeed, experiments performed at constant temperature below $T_{cc}$ (at 10 $^{o}$C) as a function of time [27] revealed that the $A_{sl} / A_f$ is strongly fluctuating, making the quantitative characterization of the cataract degree uncertain.

On the other hand, the present study demonstrated that the stretching exponent of the fast and slow decay steps – at a given temperature which corresponds to a certain degree of cataract – are really time independent parameters. We suggest that $\beta_f$ and $\beta_{sl}$ are more sensitive and reliable indicators for the *in vivo* detection of the senile cataract from the eye lens and the quantification of the lens opacity. The values of $\beta_f$ and $\beta_{sl}$ are invariant vs. time at a certains temperatures, or in other words at a certain extent of cold cataract. Therefore, the departure of the values of these parameters for unity can be regarded as reliable indicators of cataract onset.

The advantage of the present work is the extension of DLS measurements to temperatures well below the cold cataract temperature where a liquid-liquid phase separation takes place. This made it possible to extract detailed information on the dynamics of the scattering elements involved in lens opacification. The analysis showed that the dynamics of proteins bears a close resemblance to the dynamics of structural glasses where the apparent activation energy for particle diffusion increases below $T_{cc}$ indicating a highly cooperative motion. This behavior was modeled using two different methods, i.e. with the aid of two Arrhenius processes and with the use of a relation incorporating temperature dependent activation energy (VFT equation). Application of ideas developed for studying the critical dynamics of binary protein/solvent mixtures, as well as the use of the VFT equation, enabled us to estimate the spinodal temperature $T_{sp}$ of the porcine nucleus between –2 and 5 $^{o}$C, i.e. much lower than $T_{cc}$. This led to the conclusion that the protein concentration of the porcine lens nucleus is far from the critical concentration where the phase diagram shows the maximum phase separation temperature.

The microscopic mechanism underlying cold cataract is still debatable. Our data showed that lowering temperature and before reaching $T_{cc}$ there is a slight but measurable increase in the apparent protein size, which originates from changes of intermolecular interactions. Blow $T_{cc}$ the liquid-liquid phase separation takes place leading to increasing opacification. New ideas on phase separation of protein dispersions have recently appeared [28]. This effect takes place under the strong influence of viscoelastic effects owing mainly to the dynamic asymmetry (size disparity) between the components of the protein/water mixture, differentiating this effect from the phase separation in classical binary fluid mixtures [28]. Viscoelastic phase separation involves the connectivity of the slow components (protein rich phase) and the formation of a transient gel. The development of ripples observed in this study could possibly be associated with such a transit gel formation, as is well known that ripples in time correlation functions emerge inn DLS studies of gels.

Finally, the present work has demonstrated the robustness of DLS as a potential, reliable early-diagnosis tool in cataract identification. Studies in our laboratory are under way in order to investigate the cold cataract effect at different locations in the intact eye as well as to examine the effect of thermal history (pre-heating effect) of the lens on the cold cataract features.

**Acknowledgments.** Financial support of the "PENED-01 / ΕΔ-559" project is gratefully acknowledged. "PENED-01 / ΕΔ-559" project is co-funded: 75% of public financing from the European Union – European Social Fund and 25% of public financing from the Greek State – Ministry of Development – GSRT in the framework of the Operational Program "Competitiveness", Measure 8.3 – Community Support Framework 2000-2006.